\documentclass[%
 aip,
 floatfix,
rsi,%
 amsmath,amssymb,
 reprint,%
]{revtex4-1}

\usepackage{graphicx}% Include figure files
\usepackage{dcolumn}% Align table columns on decimal point
\usepackage{bm}% bold math
\usepackage{subfigure}
\usepackage{placeins}
\let\oldref\ref
\renewcommand{\ref}[1]{(\oldref{#1})}

\newcommand{\be}{\begin{equation}}
\newcommand{\ee}{\end{equation}}
\newcommand{\eenb}{\end{enumerate}}
\newcommand{\belb}{\begin{enumerate}[label= \bfseries (\alph*)]}
\newcommand{\eelb}{\end{enumerate}}

\newcommand{\dmat}{\mathbf{D}}

\newcommand{{\p}}{{\Pi}}
\newcommand{\bea}{{\begin{eqnarray}}}
\newcommand{\eea}{{\end{eqnarray}}}

\newcommand{\bK}{{\mathbf{K}}}
\newcommand{\bD}{{\mathbf{D}}}
\newcommand{\bI}{{\mathbf{I}}}
\newcommand{\bP}{{\mathbf{P}}}
\newcommand{\bR}{{\mathbf{R}}}
\newcommand{\bA}{{\mathbf{A}}}

\newcommand{\bQ}{{\mathbf{Q}}}
\newcommand{\bt}{{\mathbf{t}}}
\newcommand{\bp}{{\mathbf{p}}}
\newcommand{\bv}{{\mathbf{v}}}
\newcommand{\bu}{{\mathbf{u}}}
\newcommand{\bz}{{\mathbf{z}}}
\newcommand{\bphi}{{\mbox{\protect\boldmath$\phi$}}}
\newcommand{\bpsi}{{\mbox{\protect\boldmath$\psi$}}}
\newcommand{\bone}{{\mathbf{1}}}

\makeatletter
\newcommand*{\centerfloat}{%
  \parindent \z@
  \leftskip \z@ \@plus 1fil \@minus \textwidth
  \rightskip\leftskip
  \parfillskip \z@skip}
\makeatother

\begin{document}

\preprint{AIP/123-QED}

\title[Correlation Functions, Mean First Passage Times and the Kemeny Constant] {Correlation Functions, Mean First Passage Times and the Kemeny Constant}% Force line breaks with \\
%\thanks{}

\author{Adam Kells}
\affiliation{Department of Chemistry, Kings College London}
\author{Edina Rosta}
\affiliation{Department of Chemistry, Kings College London}
\author{Alessia Annibale$^*$}
\affiliation{Department of Mathematics, Kings College London}
\affiliation{$^*$ email: alessia.annibale@kcl.ac.uk}

\date{\today}% It is always \today, today,
             %  but any date may be explicitly specified

\begin{abstract}
Markov processes are widely used models for investigating kinetic networks. Here we collate and present a variety of results pertaining to kinetic network models, in a unified framework. The aim is to lay out explicit links between 
several important quantities commonly studied in the field, 
including mean first passage times (MFPTs), correlation functions and the Kemeny constant, and highlight some of the subtleties which are often
overlooked in the literature, while providing new insights. Results 
%following from this unified framework 
include (i) a simple physical interpretation of the Kemeny constant, (ii) a recipe to infer equilibrium distributions and 
rate matrices from measurements of MFPTs, 
potentially useful in applications, including milestoning in molecular dynamics,
and (iii) a protocol to reduce 
the dimensionality of kinetic networks, based on specific requirements that the MFPTs in the coarse-grained system should satisfy. 
It is proven that this protocol 
coincides with the one proposed by Hummer and Szabo in \cite{hummer2014optimal}
and it leads to a variational principle for the Kemeny constant. 
We hope that this study will serve as a useful reference 
for readers interested in 
theoretical aspects of kinetic networks, 
some of which underpin useful applications, including milestoning and coarse-graining. 
\end{abstract}

% \keywords{Markov state models}%Use showkeys class option if keyword
%                               %display desired
\maketitle

\section{\label{sec:intro}Introduction}
The broad applicability of Markov processes has seen them used in a wide variety of fields. This has resulted in many aspects of the theory being derived on multiple occasions in diverse ways. In this study we aim to 
present a unified framework that links several results in the literature and provide some novel insights, including a simple interpretation of Kemeny constants, a recipe to construct 
rate matrices from MFPTs measuremens, and 
the definition of computationally efficient
protocols to reduce the dimensionality of Markov State models. The manuscript is organised as follows. In Sec. \ref{sec:Theory} we review spectral properties of transition and rate matrices 
in Markovian dynamics
and provide explicit expressions for the mean 
first passage times (MFPTs) 
%of discrete and continuous time Markovian dynamics 
in terms 
of their eigenvalues and eigenvectors. In Sec. \ref{sec:results}  
we derive formulae for the MFPTs in terms of rate matrices and 
correlation functions and 
give a physical interpretation for the Kemeny constant, as well as 
a recipe to construct rate matrices from measurements of 
MFPTs, a problem with relevance in milestoning techniques 
\cite{milestoning_Elber04, milestoning_Vanden08, milestoning_Vanden2018, milestoning_Elber17}. 
Taking advantage 
of these relations, we propose a protocol to reduce the dimensionality 
of kinetic networks, based on the requirement that 
a certain relation between the MFPTs of the original and the 
coarse-grained system is satisfied. We show that this protocol coincides with 
the coarse-graining proposed recently by Hummer and Szabo in \cite{hummer2014optimal}, 
and it leads to a variational principle for the 
Kemeny constant, which can be potentially useful to optimise the coarse-graining.

\section{Theory}
\label{sec:Theory}
\subsection{Markov Chains}
A kinetic network consists of $n$ discrete states labelled $i=\{1,...,n\}$. Each discrete state has a time dependent probability to be occupied $p_i(t)$. The evolution of these probabilities, in continuous time, is governed by the rate at which the system moves between different states. The rate $k_{ji}$ of transition from state $i$ to state $j$ is given by 
\begin{equation}
    k_{ji}=\lim_{\tau\rightarrow 0}\frac{P(j,t+\tau|i,t)}{\tau},
\end{equation}
where $P(j,t+\tau|i,t)$ is the probability to make the transition in a small interval of time $\tau$.
The time-evolution of the probability of state occupation is given by the master equation
\begin{equation}
    \frac{dp_i(t)}{dt}=\sum_{j\neq i}\bigg[k_{ij}p_j(t) - k_{ji}p_i(t)\bigg], 
    \end{equation}
which can be written in matrix notation
\begin{equation}\label{eq:rate}
    \frac{d\bp}{dt}=\bK\bp
\end{equation}
using the fact that the diagonal elements of the rate matrix $\bK$ are necessarily given by $k_{ii}=-\sum_{j}k_{ji}$ for conservation of probability.
If $\bK$ has a complete set of eigenvectors, equation (\ref{eq:rate}) is solved by
\begin{equation} \label{eq:sol_ME}
    \bp(t) = e^{\bK t}\bp(0),
\end{equation}
where the so-called propagator $e^{\bK t}$ is a matrix which evolves the probability distribution at one time to a new distribution at a time $t$ later. 

In discrete time $t=\ell\tau$, where moves between states happen at multiples $\ell=1,2,\ldots$ of a given time interval $\tau$, 
one defines the transition matrix $\bQ(\tau)=e^{\bK \tau}$, whose elements 
give the transition probability over a single time step, for any pair of states. The probability vector at the $\ell$-th time step can then be found as 
\begin{equation}
    \bp(\ell) = [\bQ(\tau)]^\ell\bp(0).
\end{equation}
We will draw particular attention to the distinction between continuous and discrete time dynamics, when deriving MFPTs expressions.

\subsection{Eigenvalues and Eigenvectors}

The rate matrix can be spectrally decomposed and represented in terms of its eigenvalues $\{\lambda_\ell\}_{\ell=1}^n$ and left and right eigenvectors, $\{\bphi^{(\ell)}\}_{\ell=1}^n$ and $\{\bpsi^{(\ell)}\}_{\ell=1}^n$, respectively
\begin{equation}
\label{eq:spectral_K}
    \bK = \sum_{\ell=1}^n\lambda_\ell \bpsi^{(\ell)}\bphi^{(\ell)}.
\end{equation}
We will focus on systems satisfying detailed balance, where 
eigenvalues are real.
The largest eigenvalue of $\bK$ is $0$ and so all other eigenvalues are negative. They are usually indexed in descending order
\begin{equation}
    0=\lambda_1\geq \lambda_2 \geq ... \geq \lambda_N.
\end{equation}
The corresponding eigenvectors are indexed in the same manner. The right eigenvector corresponding to the zero eigenvalue $\bpsi^{(1)}$ is known as the stationary probability (or, for reversible dynamics, equilibrium probability)
$\bp^{\rm eq}$ with elements $p^{\rm eq}_i$. The corresponding 
left eigenvector $\bphi^{(1)}$ is the $n$-dimensional row vector with all the 
components equal to $1$, $\bone_n^T$. 

It can be shown that the elements of the left and right eigenvectors are related by the equilibrium probability
\begin{equation}
\label{eq:left_right}
    \psi_i^{(\ell)}=\phi^{(\ell)}_ip^{\rm eq}_i
\end{equation}
and $\sum_i \psi_i^{(\ell)}=0$ for $\ell >1$.
Hence, left and right 
eigenvectors associated to non-zero eigenvalues 
will have positive and negative entries. 
These contain useful kinetic information, as they 
are related to relaxation processes. 

This link can be seen by using the spectral decomposition (\ref{eq:spectral_K}) in equation (\ref{eq:sol_ME})
and singling out the contribution from $\ell=1$
\be
p_i(t)-p_i^{\rm eq}=\sum_{\ell\geq 2}^n e^{-|\lambda_\ell| t} \psi_i^{(\ell)}\bphi^{(\ell)}\cdot\bp(0),
\label{eq:dominant_contribution}
\ee
where we have used $\psi_i^{(1)}=p_i^{\rm eq}$, $\phi_j^{(1)}=1~\forall~j$, $\sum_j p_j(0)=1$ and $\lambda_\ell<0 ~\forall~\ell\geq 2$. 
%As $|\lambda_\ell|>|\lambda_2|~\forall~\ell>2$, 
For large time, the RHS of (\ref{eq:dominant_contribution}) is dominated by the first term in the sum, %showing 
so the probability distribution will tend towards the equilibrium distribution with a timescale given by $\tau_2=1/|\lambda_2|$ (often called 
the relaxation time). 
The other timescales, are each given by the inverse of the magnitude of the corresponding eigenvalue
\begin{equation}
 \tau_\ell=1/|\lambda_\ell|
\end{equation}
%
%%%%%%%%%%%
and can be interpreted as the time with which the rate matrix moves probability density between the oppositely signed regions of the corresponding eigenvector. 
This can be seen by considering the evolution of the scalar product between the
time-dependent probability and the different eigenvectors 
\begin{equation}
   \bphi^{(s)}\cdot\bp(t)= e^{-|\lambda_s| t} 
   \bphi^{(s)}\cdot\bp(0).
   \label{eq:scalar_pro}
\end{equation}
%obtained by using the spectral decomposition \ref{eq:spectral_K} in equation %\ref{eq:sol_ME} and projecting onto left eigenvector $\bphi^{(s)}$. 
Each scalar product vanishes on a timescale set by the inverse eigenvalue, indicating that the probability mass becomes distributed evenly across positive and negative entries of the eigenvector $\bphi^{(s)}$, on the timescale $1/|\lambda_s|$. 

\subsection{Correlation Functions}
The correlation function between two observables $\theta_i$ and $\theta_j$ at a lagtime $\tau$ is 
given by 
\begin{equation}
    C_{ji}(\tau,t) = \langle \theta_j(t+\tau) \theta_i(t) \rangle -\langle \theta_j(t+\tau)\rangle\langle \theta_i(t) \rangle
    \label{eq:correlation}
    \end{equation}
Defining $\theta_i(t)$ as the indicator function which takes value $1$ when the system is in state $i$ at time $t$ and 0 otherwise, the first term of (\ref{eq:correlation}) gives the joint probability that the 
system is in state $i$ at time $t$ and in state $j$ at a time $\tau$ later
\begin{eqnarray}
C_{ji}(\tau,t)&=&P(j,t+\tau;i,t)-p_j(t+\tau)p_i(t)
\nonumber\\
&=&[P(j,t+\tau|i,t)-p_j(t+\tau)]p_i(t)
\label{eq:dynamical}
\end{eqnarray}
where the conditional probability $P(j,t+\tau|i,t)$ is given by the $ji$'th entry 
of the propagator matrix, and depends only on the lagtime $\tau$, i.e. $P(j,t+\tau|i,t)=[e^{\bK \tau}]_{ji}=P(j,\tau|i,0)$.
If the system is in equilibrium, where one-time quantities are time-independent, 
the correlation function becomes a function of only the lagtime
\begin{equation}
    C^{\rm eq}_{ji}(\tau)=[e^{\bK \tau}]_{ji}p_i^{\rm eq}-p_j^{\rm eq}p_i^{\rm eq}.
    \label{eq:Ceq}
\end{equation}
In many practical situations, one averages (\ref{eq:dynamical}) over the earlier time $t$, with the expectation that if the system is ergodic (i.e. a sufficiently long trajectory will sample all states with equilibrium probability) the resulting time average equates the equilibrium correlator
\begin{equation}
    \overline{C_{ji}(\tau,t)}=\lim_{T\rightarrow \infty}\frac{1}{T}\int_0^T dt\, C_{ji}(\tau,t)\equiv C_{ji}^{\rm eq}(\tau).
\end{equation}
Repeating the same steps that led to equation (\ref{eq:dominant_contribution}), the equilibrium correlator 
(\ref{eq:Ceq}) can be written as 
a superposition of exponential functions 
\begin{equation}
    C^{\rm eq}_{ji}(\tau)=\sum_{\ell\geq 2} e^{-|\lambda_\ell| \tau} \psi^{(\ell)}_j\phi^{(\ell)}_i p_i^{\rm eq} 
    \label{eq:spectral_C}
\end{equation}
decaying to zero at large lagtime.
The area underneath the correlator, then serves as 
a measure of how quickly an initial probability distribution will tend to the equilibrium probability, and it
%. This is known as the 'decorrelation time' 
can be expressed
%concisely by spectral decomposition of the propagator 
as a weighted sum of the timescales in the system
\begin{equation}
\label{eq:corr_int_spec}
    \int_0^{\infty} C^{\rm eq}_{ji}(\tau)d\tau
    = \sum_{\ell \geq 2} \frac{1}{|\lambda_\ell|}\psi^{(\ell)}_j\phi^{(\ell)}_i p_i^{\rm eq}=  \sum_{\ell\geq 2} \tau_\ell\, \psi^{(\ell)}_j\psi^{(\ell)}_i
\end{equation}
where we have also used (\ref{eq:left_right}).
One final observation that will be useful in this study is that 
the above quantities can be rewritten as
\begin{equation}\label{eq:corr_inv}
    \int_0^{\infty} C^{\rm eq}_{ji}(\tau)d\tau = (\bp^{\rm eq}\mathbf{1}_n^T-\bK)^{-1}_{ji}p^{\rm eq}_i -p_j^{\rm eq}
    p_i^{\rm eq},
\end{equation}
where we have used $\bp^{\rm eq}=\bpsi^{(1)}$,  $\bone_n^T=\bphi^{(1)}$ and $(\bpsi^{(1)}\bphi^{(1)}-\bK)^{-1}=
\bpsi^{(1)}\bphi^{(1)}-\sum_{\ell\geq 2}\lambda_\ell^{-1} \bpsi^{(\ell)}\bphi^{(\ell)}
$.

\subsection{Mean First Passage Time}

Next, %with the fundamental theory of Markov chains, 
%we can examine how to calculate MFPTs. We want to write down 
we derive an expression for MFPTs, i.e. the expected time it takes to the system to first reach a state $j$ given its current state is $i$, $t_{ji}$, within the fundamental theory of Markov processes. We will consider the discrete and continuous time cases separately to highlight the subtle theoretical difference between the two cases. 

\subsubsection{Discrete Time}
First we consider the case where the system can make transitions at discrete intervals, without loss of generality we define our units of time such that this time interval is $1$. This system is defined by a transition matrix $\bQ$, such that $\sum_j Q_{ji}=1~\forall~i$, which has eigenvalues 
$1=\lambda_1'\geq \lambda_2'\geq\ldots\geq\lambda_N'$ and 
eigenvectors as for the rate matrix $\bK$. 

We will use a prime index to denote quantities in discrete time dynamics that differ from their analogues in continuous time dynamics, for which we will use the same symbols without the prime.
Accordingly, we will denote with $t_{ji}'$ the mean number of {\it time steps} that it takes to the system to first reach $j$ from $i$, in discrete time dynamics, 
whereas the corresponding quantity in continuous time dynamics will be denoted with $t_{ji}$, and will measure the mean  
{\it time} for the first visit to $j$, from $i$, to occur.

When the system starts in state $i$, it can either move to $j$ directly (i.e. in one time step), with probability $Q_{ji}$, or 
transition to some other state $k$ with probability $Q_{ki}$ (in one time step) and then move to $j$ in a time of $t'_{jk}$, ($t'_{jk}\!+\!1$ in total), leading to the recursion 
\begin{equation}\label{eq:mfpt_disc1}
    t'_{ji}=Q_{ji}+\sum_{k\neq j}(t'_{jk}+1)Q_{ki}=1+\sum_{k\neq j}t'_{jk} Q_{ki}.
\end{equation}
We can rewrite (\ref{eq:mfpt_disc1}) as
\begin{equation}\label{eq:mfpt_disc2}
    \sum_{k}t'_{jk}(\delta_{ki}-Q_{ki})=1-Q_{ji}t'_{jj}
\end{equation}
where $\delta_{ki}$ is the Kronecker delta, that leads to the more convenient matrix form
\be
{\bt'_j}^T (\bI-\bQ)=(1-Q_{j1}t'_{jj},\ldots,1-Q_{jN}t'_{jj})
\label{eq:tj}
\ee
where we have defined ${\bt'}_j^T=(t'_{j1},\ldots,t'_{jN})$ as the row 
vector with the MFPTs to $j$ as components. 

If $\bQ$ has a complete set of orthonormal eigenvectors (which is guaranteed if detailed balance is satisfied), one can express ${\bt'}_j^T$ as a 
linear combination of the (left) eigenvectors of $\bQ$, for certain coefficients $a_{nm}$ to be determined a posteriori
\be
\label{eq:lin_comb}
{\bt'_j}^T=\sum_\ell a_{j\ell}\bphi^{(\ell)}.
\ee
Inserting in equation (\ref{eq:tj}) gives the vector equation
\be
\sum_\ell a_{j\ell} (1-\lambda'_\ell)\bphi^{(\ell)}=(1-Q_{j1}t'_{jj},\ldots,1-Q_{jN}t'_{jj}).
\ee
Next we consider the equation for the component $r$ 
\be
\sum_\ell a_{j\ell} (1-\lambda'_\ell)\phi_r^{(\ell)}=1-Q_{jr}t'_{jj}.
\ee
Multiplying left and right hand sides times $\psi_r^{(s)}$ and summing over $r$ gives
\be
\sum_{\ell>1} a_{j\ell} (1-\lambda'_\ell)\delta_{\ell s}=\delta_{s 1}-\lambda_s \psi_j^{(s)}t'_{jj}
\label{eq:general_s}
\ee
where we have used that $\bpsi^{(s)}$ is the right eigenvector of $\bQ$ associated to eigenvalue $\lambda'_s$, and the properties of the eigenvectors of the matrix $\bQ$, $\sum_r \psi_r^{(s)}=\delta_{s1}$, and $\sum_r \phi_r^{(\ell)}\psi_r^{(\ell)}=\delta_{\ell s}$.
Equation (\ref{eq:general_s}) yields for $s=1$ 
\be
t'_{jj}=\frac{1}{p^{\rm eq}_j}
\label{eq:return}
\ee
This quantity is greater than 
or equal to one, with equality holding for 
$p_j^{\rm eq}=1$, and it can be interpreted as the expected number of time steps it takes to the system to first hit state $j$, {\it after} its release from state $j$ itself, also known as the "recurrence time" or Kac's lemma \cite{Kac1947}.   
At this point it should be noted that some studies in the literature set this quantity to zero as a 'convention'. The analysis above shows that, in the discrete time formulation of MFPTs, the convention (\ref{eq:return}) should be used. 
For $s>1$, using (\ref{eq:return}) one gets from equation (\ref{eq:general_s})
\be
\label{eq:ajs}
a_{js}=-\frac{1}{p^{\rm eq}_j}\frac{\lambda'_s}{1-\lambda'_s} \psi_j^{(s)}.
\ee
Singling out the contribution from $a_{j1}$ in (\ref{eq:lin_comb})
\be
\bt'_j=a_{j1}\bphi^{(1)}+\sum_{\ell>1}a_{j\ell}\bphi^{(\ell)}
\ee
using $\phi_k^{(1)}=1~\forall~k$ and (\ref{eq:ajs}), we get
\be
t'_{jk}=a_{j1}-\frac{1}{p^{\rm eq}_j}\sum_{\ell>1}\frac{\lambda'_\ell}{1-\lambda'_\ell}\psi_j^{(\ell)}\phi_k^{(\ell)}
\label{eq:tjk}
\ee
where $a_{j1}$ can be determined by setting $j=k$ in the above and using (\ref{eq:return})
\be
a_{j1}=\frac{1}{p^{\rm eq}_j}
\left(1+\sum_{\ell>1}\frac{\lambda'_\ell}{1-\lambda'_\ell}\phi_j^{(\ell)}\psi_j^{(\ell)}\right).
\ee
Substituting in (\ref{eq:tjk}), we finally obtain an explicit relation
for the MFPTs in terms of the eigenvalues and eigenvectors 
of the transition matrix
\be
t'_{jk}=\frac{1}{p^{\rm eq}_j}
\left[1+\sum_{\ell>1}\frac{\lambda'_\ell}{1-\lambda'_\ell}\psi_j^{(\ell)}
(\phi_j^{(\ell)}-\phi_k^{(\ell)})\right].
\label{eq:mfpt_discrete}
\ee
This formula, also derived in \cite{mfpt_eigenvec1, mfpt_eigenvec2}, will serve as a starting point to derive a number of useful relations in the following sections.

\subsubsection{Continuous Time}
Next we consider how these results differ when our system is described by a continuous time rate matrix $\bK$ instead of a discrete time transition probability matrix. 
Results for continuous time dynamics can be derived by setting the time step to $\tau$ in the discrete time dynamics, and taking the limit $\tau\to 0$ at the end.  
For small but finite $\tau$, the transition matrix $\bQ$
can be written as $e^{\bK \tau}$, and its eigenvalues are given by 
$\lambda_\ell'=e^{\lambda_\ell \tau}$. 
%We will take a time unit of $\tau$ and consider the limit to zero at the end. 
Defining $t_{ji}=t_{ji}'\tau$ as the mean first time from $i$ to $j$, and using the same logic as in (\ref{eq:mfpt_disc1}), we can write a similar recursion
\begin{equation}\label{eq:mfpt_cont1}
    t_{ji}=[e^{\bK \tau}]_{ji}\tau+\sum_{k\neq j}[e^{\bK \tau}]_{ki}(t_{jk}+\tau)=\tau+\sum_{k\neq j}[e^{\bK \tau}]_{ki}t_{jk},
\end{equation}
that can be rearranged as in equation (\ref{eq:mfpt_cont2}),
\begin{equation}\label{eq:mfpt_cont2}
    \sum_{k}(\delta_{ki}-[e^{\bK \tau}]_{ki})t'_{jk}=1-[e^{\bK \tau}]_{ji}t'_{jj}.
\end{equation}
Following the same steps that led to (\ref{eq:mfpt_discrete}) we can arrive at 
\begin{equation}\label{eq:mfpt_cont3}
    t'_{ji}=\frac{1}{p^{\rm eq}_j}\bigg[1 + \sum_{\ell>1}\frac{e^{\lambda_\ell \tau}}{1-e^{\lambda_\ell\tau}} \psi_{j}^{(\ell)}(\phi_j^{(\ell)} - \phi_i^{(\ell)})\bigg]
\end{equation}
Finally, using $t_{ji}=t'_{ji}\tau$ and taking the limit $\tau\to 0$, gives a formula for the MFPTs in continuous time dynamics, in terms of eigenvalues and eigenvectors of the rate matrix
\begin{equation}\label{eq:mfpt_cont4}
    t_{ji}=\frac{1}{p^{\rm eq}_j} \sum_{\ell>1}\frac{1}{|\lambda_\ell|} \psi_{j}^{(\ell)}(\phi_j^{(\ell)} - \phi_i^{(\ell)}).
\end{equation}
Note that in contrast to the discrete time result (\ref{eq:return}), 
in continuous time dynamics, equation (\ref{eq:mfpt_cont4}) implies 
\begin{equation}
    t_{jj}=0,
    \label{eq:return_ct}
    \end{equation}
which is intuitively understood, as here there is no time step 
to wait to return to the state. 

As an aside, we observe that expanding (\ref{eq:mfpt_cont2}) for small $\tau$
as in equation (\ref{eq:limit})
\begin{equation}
    -\tau \sum_k t'_{jk}K_{ki}=1-(\delta_{ji}+\tau K_{ji})t'_{jj}
    \label{eq:limit}
\end{equation}
using (\ref{eq:return}), $t_{ji}=\tau t'_{ji}$ and 
{\it then} letting $\tau\to 0$, gives
\begin{equation}
    \bt \bK=-\bone_n \bone_n^T+\bD_n^{-1},
    \label{eq:correct}
\end{equation}
where $\bD_n$ is an $n\times n$ diagonal matrix 
with $\bp^{\rm eq}$ on the 
diagonal.
%Note that \ref{eq:correct} was obtained by setting $t'_{jj}=1/p_j^{\rm eq}$ before taking the limit $\tau\to 0$. 
Note that the order in which these operations are executed matters, as 
$t'_{jj}$ and $t_{jk}$ (with $j\neq k$) should remain finite as $\tau$ is sent to zero. Taking the limit naively,
leads to the expression given in (\ref{eq:wrong}), with $\bt_j^T=(t_{j1},\ldots,t_{jN})$
\begin{equation}
    \bt_j^T \bK = -\bone_n^T,
    \label{eq:wrong}
\end{equation}
which is sometimes reported in the literature. This is equivalent to
$\bt \bK=-\bone_n \bone_n^T$, thus it differs from (\ref{eq:correct}) 
for the diagonal terms. It is easy to show that (\ref{eq:correct}) 
is correct, while (\ref{eq:wrong}) is not, e.g. by multiplying both expressions 
times $\bp^{\rm eq}$ from right and using $\bK\bp^{\rm eq}=0$, 
$\bone_n^T \bp^{\rm eq}=1$ and $\bD_n^{-1}\bp^{\rm eq}=\bone_n$.

Finally we note that, although (\ref{eq:correct}) provides a correct expression for the MFPTs, $\bK$ is not directly invertible due the presence of zero 
eigenvalues, hence MFPTs are more easily computed from relations 
that we will derive in the next sections, 
which directly follow from 
(\ref{eq:mfpt_cont4}).

\subsection{Kemeny Constant}
Starting with equation (\ref{eq:mfpt_cont4}) we can examine the quantity $\sum_{j}p_j^{\rm eq}t_{ji}$ and make use of $\sum_j\psi_j^{(\ell)}=\delta_{\ell,1}$ and $\sum_j\phi^{(\ell)}_j\psi_j^{(\ell)}=1$ for all $\ell$, to get
\begin{eqnarray}
    \sum_{j}p_j^{\rm eq}t_{ji}&=&\sum_j \sum_{\ell>1}\frac{1}{|\lambda_\ell|} \psi_{j}^{(\ell)}(\phi_j^{(\ell)} - \phi_i^{(\ell)})=\sum_{\ell>1}\frac{1}{|\lambda_\ell|}
    \nonumber\\
    &=&\sum_{\ell>1}\tau_\ell \equiv \zeta.
    \label{eq:zeta}
\end{eqnarray}
This result is known as the Kemeny constant \cite{KemenySnell1960, doyle2009kemeny, Kemeny_hunter14}
and is remarkable as it relates a weighted sum of MFPTs starting from some state $i$ to a sum over relaxation timescales (which is independent of the particular choice of $i$).

The corresponding quantity in discrete time dynamics is obtained 
summing (\ref{eq:mfpt_discrete}) over $j$
\begin{eqnarray}
\sum_j p_j^{\rm eq} t_{jk}'-N&=&
\sum_{\ell>1}\frac{\lambda_\ell'}{1-\lambda_\ell'}
(1-\delta_{\ell,1})
=
\sum_{\ell>1}\left(\frac{1}{1-\lambda_\ell'}-1\right)
\nonumber\\
&=&\sum_{\ell>1}\frac{1}{1-\lambda_\ell'}-(N-1)
\end{eqnarray}
which, simplifies, using (\ref{eq:return}), to
\be
\sum_{j(\neq k)}p_j^{\rm eq}t'_{jk}=\sum_{\ell>1}\frac{1}{1-\lambda_\ell'}
\equiv \zeta'.
\label{eq:zeta_disc}
\ee
We conclude this section by noting that there have been several studies 
focusing on the Kemeny constant in the field of graph theory and networks 
science. Loosely speaking, a low Kemeny constant 
means that the time to travel between 
states is on average small, so this is interpreted to mean that the graph 
is well-connected \cite{kemeny_nets}. The Kemeny constant can be used to calculate 
the Kirchoff index of a graph \cite{palacios} and it has been 
proposed as an objective function to optimize in graph clustering 
algorithms \cite{kemeny_nets2}.

\section{Results}
\label{sec:results}
With the theory laid out, we are now equipped to make some observations about how these quantities relate. In particular we will show two main results:

\begin{itemize}
    \item A description of how MFPTs and Kemeny constant are related to rate matrices and correlation functions. This will lead to a simple interpretation of the Kemeny constant and to a recipe for
    reconstructing rate matrices from MFPTs measurements, which may be helpful in milestoning\cite{milestoning_Elber04,milestoning_Vanden08,milestoning_Elber17,milestoning_Vanden2018,milestoning_Szabo19} and transition path sampling \cite{TPS1, dellago_chandler_TS_98, TPS2, TPS3, TPS4}.
    \item An example of how this unified framework can be applied to derive a coarse grained rate matrix which ensures that the MFPTs of the high dimensional and low dimensional systems are the same.
\end{itemize}
From now on, we will focus on continuous time dynamics, as much of the focus on MFPTs in the literature is for discrete time dynamics.

\subsection{Linking MFPTs and Kemeny Constants To Correlation Functions}

In this section, we provide expressions for MFPTs in terms 
of rate matrices and correlation functions and provide a physical interpretation 
for Kemeny constants. 
We start by adding and subtracting $\psi_j^{(1)}$ from equation 
(\ref{eq:mfpt_cont4}), using $\phi_i^{(1)}=1~\forall~i$ and 
$|\lambda_\ell|=-\lambda_\ell~\forall~\ell>1$
\begin{eqnarray}
     t_{ji}&=&\frac{1}{p_j^{\rm eq}}\left[\psi_{j}^{(1)}\phi_j^{(1)}-\sum_{\ell>1}\frac{1}{\lambda_\ell} \psi_{j}^{(\ell)}\phi_j^{(\ell)}
    -\psi_{j}^{(1)}\phi_i^{(1)}\right.
    \nonumber\\
    &&+\left.\sum_{\ell>1}\frac{1}{\lambda_\ell}\psi_{j}^{(\ell)}\phi_i^{(\ell)}\right]
     \end{eqnarray}
     to reformulate the 
expression for the MFPTs in terms of matrix elements
     \begin{eqnarray}
t_{ji}= \frac{1}{p_j^{\rm eq}}\bigg[ (\bp^{\rm eq}\mathbf{1}_n^T-\bK)^{-1}_{jj}-(\bp^{\rm eq}\mathbf{1}_n^T-\bK)^{-1}_{ji} \bigg]
     \label{eq:t_K}
\end{eqnarray}
where we have used $\bp^{\rm eq}=\bpsi^{(1)}$ and $\bone_n^T=\bphi^{(1)}$.
This gives an explicit formula for MFPTs 
in continuous time dynamics, in terms of rate matrices, 
which complements 
similar results available in the literature for discrete time 
dynamics \cite{Grinstead2003}, formulated in terms of the so-called 'fundamental matrix' $(\bp\bone_n^T+\bI-\bQ)^{-1}$.
Now using equation (\ref{eq:corr_inv}), one can provide yet another expression for MFPTs, in terms of time-integrated correlation functions
\begin{equation}
    t_{ji}=\frac{1}{p_j^{\rm eq}}\left[\frac{\int_{0}^{\infty}C^{\rm eq}_{jj}(\tau)d\tau}{p_j^{\rm eq}}-\frac{\int_{0}^{\infty}C^{\rm eq}_{ji}(\tau)d\tau}{p_i^{\rm eq}}\right]
    \label{eq:MFPT_C}
\end{equation}
which is appealing as it does not require the inversion of a high dimensional matrix, in the same way as (\ref{eq:t_K}) does. The Kemeny constant follows as
\begin{equation}
    \zeta=\sum_jp_j^{eq}t_{ji}=\sum_j \bigg[\frac{\int_{0}^{\infty}C^{\rm eq}_{jj}(\tau)d\tau}{p_j^{\rm eq}}-\frac{\int_{0}^{\infty}C^{\rm eq}_{ji}(\tau)d\tau}{p_i^{\rm eq}}\bigg].
    \label{eq:zeta_C}
\end{equation}
Since $C_{ji}^{\rm eq}(\tau)/p_i^{\rm eq}=P(j,\tau|i,0)-p_j^{\rm eq}$
and $\sum_j P(j,\tau|i,0)=1 ~\forall~\tau$, swapping sums with integrals 
in (\ref{eq:zeta_C}), which is valid for finite state space,
it becomes clear that the second term on the 
RHS vanishes, giving
\begin{eqnarray}
    \zeta&=&\sum_j\frac{\int_{0}^{\infty} C_{jj}^{\rm eq}(\tau)d\tau}{p_j^{\rm eq}}
       \label{eq:interpretation_zeta}
    \\
    &=&\sum_j \int_{0}^{\infty} [P(j,\tau|j,0)-p_j^{\rm}]d\tau
    \label{eq:interpretation_zeta2}
\end{eqnarray}
The first term in the square brackets measures the fraction of trajectories that are in $j$ at time $\tau$, out of those that 
start in $j$ at time $0$. The second term measures the fraction of trajectories that are in $j$ at a given time 
$\tau$, out of all the trajectories. 
Equation (\ref{eq:interpretation_zeta2}) reveals that Kemeny constant can be regarded as the time-integrated difference between the conditional and the a priori probability to be in any given state, as similarly pointed out in \cite{Bini}. Furthermore, equation (\ref{eq:interpretation_zeta}) shows that $\zeta$ can be written as the trace of a matrix, that is known as the 'deviation matrix' \cite{dev_matrix, Bini}.

A more convenient writing of (\ref{eq:interpretation_zeta2}), 
which avoids its formulation in terms 
of the (finite) difference between two divergent 
integrals,  
can be obtained by introducing the decorrelation time of a state $j$
\begin{equation}
    T_j=\int_0^{\infty}\frac{C_{jj}^{\rm eq}(\tau)}{C_{jj}^{\rm eq}(0)}d\tau,
\end{equation}
as the area underneath the normalised autocorrelation functions $\hat C_{jj}^{\rm eq}(\tau)=C_{jj}^{\rm eq}(\tau)/C_{jj}^{\rm eq}(0)$. The latter takes values $1$ for 
$\tau=0$ and zero for $\tau\to \infty$, and it decays as a multi-exponential, thus yielding a convergent integral. Using $C_{jj}^{\rm eq}(0)=p_j^{\rm eq}(1-p_j^{\rm eq})$, one can express the Kemeny constant 
%in terms of the decorrelation times of the individual states, 
as in (\ref{eq:weight_sum})
\begin{equation}
\label{eq:weight_sum}
    \zeta=\sum_j T_j (1-p_j^{\rm eq}).
\end{equation}
This leads to a simple interpretation of the 
Kemeny constant,
as a weighted sum of the decorrelation times of the individual states. 
Here, $1-p_j^{\rm eq}$ can be thought of as the difference between the maximum value, $1$, and the minimum value, $p_j^{\rm eq}$, of the conditional probability $P(j,\tau|j,0)$, (attained at $\tau=0$ and $\tau=\infty$ respectively), while $T_j$ measures how fast $P(j,\tau|j,0)$ decays from the former to the latter value.
Note that for systems with a large number of states $n$ and broad equilibrium distribution, one is normally interested in, individual state probabilities are small, i.e. $p_j^{\rm eq}\ll 1~\forall~j$, hence 
\begin{equation}
    \zeta\simeq \sum_{j=1}^n T_j,\quad n\gg 1
    \end{equation}
Finally, we note that combining (\ref{eq:weight_sum}) and
(\ref{eq:zeta}) provides an intriguing chain of relations for MFPTs, decorrelation times and relaxation times 
\begin{equation}
    \sum_{j=1}^n p_j^{\rm eq}t_{ji}=\sum_{j=1}^n T_j (1-p_j^{\rm eq})=
    \sum_{\ell=2}^n\tau_\ell.
\end{equation}

\iffalse
Finally we note that by using (\ref{eq:corr_int_spec})
decorrelation times can be written as a weighted sum of relaxation 
times
\begin{equation}
    T_j=\sum_{\ell>1}\tau_\ell \frac{(\psi_j^{(\ell)})^2}{\sum_{r>1}(\psi_j^{(r)})^2}
\end{equation}
\fi

\subsection{Constructing Rate Matrices from MFPTs}
With an explicit expression for MFPTs in terms of rate matrices, we can now invert this expression, to obtain a recipe for 
constructing rate matrices with given MFPTs. 
Upon defining $\bz$ as the vector with components 
$z_j=[(\bp^{\rm eq} \bone_n^T-\bK)^{-1}]_{jj}$, 
we can write (\ref{eq:t_K}) in 
matrix form 
\begin{equation}
    \bD_n \bt=\bz\bone_n^T-(\bp^{\rm eq} \bone_n^T-\bK)^{-1}
    \label{eq:D_K}
\end{equation}
Rearranging, we obtain
\be
\bK=\bp^{\rm eq}\bone_n^T -(\bz\bone_n^T-\bD_n \bt)^{-1},
\label{eq:K_inf}
\ee
where $\bz$ can be expressed in terms of $\bt$ by demanding $\bK\bp^{\rm eq}=0$
\begin{equation}
    \bz=\bp^{\rm eq}+\bD_n\bt\bp^{\rm eq}.
\end{equation}
Substituting into (\ref{eq:K_inf}) then gives
\be
\bK=\bp^{\rm eq}\bone_n^T -[\bp^{\rm eq}\bone_n^T-\bD_n \bt(\bI-\bp^{\rm eq}\bone_n^T)]^{-1}.
\label{eq:K_inferred}
\ee
It is easy to show that (\ref{eq:K_inferred}) also satisfies $\bone_n^T \bK=0$, by noting that $\bone_n^T \bD_n=[\bp^{\rm eq}]^T$ and  
\be
[\bp^{\rm eq}]^T \bt=\zeta \bone_n^T,
\label{eq:zeta_def}
\ee
which is implied by the definition of Kemeny constant (\ref{eq:zeta}). Equation (\ref{eq:zeta_def}) also shows that the equilibrium distribution can be fully determined 
from the matrix of MFPTs, as $[\bp^{\rm eq}]^T=\zeta \bone_n^T \bt^{-1}$ where $\zeta$ follows from the normalization of $\bp^{\rm eq}$, as $\zeta=1/(\bone_n^T \bt^{-1}\cdot \bone_n)$, so 
\begin{equation}\label{eq:eq_t}
    [\bp^{\rm eq}]^T=\frac{\bone_n^T \bt^{-1}}{\bone_n^T \bt^{-1}\cdot \bone_n}. 
    \end{equation}
By using (\ref{eq:zeta_def}) and the Sherman-Morrison formula, as 
shown in the Appendix, 
equation (\ref{eq:K_inferred}) can be simplified to obtain 
\begin{equation}
    \bK=\bt^{-1}(\bD_n^{-1}-\bone_n \bone_n^T),
    \label{eq:K_inf2}
\end{equation}
which can also be derived from (\ref{eq:correct}).
Since $\bD_n$ follows directly from $\bp^{\rm eq}$, 
equations (\ref{eq:eq_t}) and (\ref{eq:K_inf2}) show that 
$\bp^{\rm eq}$ and $\bK$ can be both computed by inverting a 
single matrix (i.e. $\bt$).

These equations then 
give a recipe to infer the equilibrium probability %$\bp^{\rm eq}$ and the rate matrix %$\bK$ 
of a system with $n$ states, 
from the {\it sole} 
observation of MFPTs between pairs of states.  
This may be useful in 
practical situations where 
information about MFPTs is readily available, whereas 
information about the rate matrix and the equilibrium distribution is not. 

We note that in Markov processes with ordered states, reflecting boundary conditions, and transitions only occurring 
between adjacent states, one has, for any pair of states $i<j$, 
$t_{ij}=\sum_{k=i}^{j-1}t_{k,k+1}$. 
Hence, the full matrix $\bt$ can be 
determined from the knowledge of only MFPTs between adjacent 
states, $t_{k,k\pm 1}, \forall~k$. Equations (\ref{eq:eq_t}) and (\ref{eq:K_inf2}) can then be used to reconstruct the full equilibrium distribution and rate matrix, from the observation of MFPTs between adjacent states, which can be computed efficiently, e.g. via the trajectory coloring procedure introduced in \cite{coloring_Vanden2010,coloring_VandenVenturoli09}. This can be useful in milestoning procedures, aimed at inferring the full kinetics of a system from the observation of many short trajectories, between 
adjacent states (milestones). 

We note that for milestoning on one-dimensional potentials,
recipes to construct rate matrices have been given in terms 
of MFPTs {\it and} committor probabilities for adjacent milestones
\cite{milestoning_Szabo19, milestoning_Vanden08}.
Equation (\ref{eq:K_inf2}), equipped with (\ref{eq:eq_t}), 
provides an alternative route 
%to compute the rate matrix, 
which does not require to estimate committor probabilities. 
The above framework 
provides an intuitive explanation for the observed accuracy of  
milestoning techniques, when applied to one-dimensional 
Smolochowski processes, 
in predicting the full distribution of MFPTs, by using rate 
matrices constructed 
from MFPTs between adjacent milestones \cite{milestoning_Szabo19, milestoning_Vanden08}: for these processes, MFPTs between adjacent states are sufficient to construct the whole MFPTs matrix, 
which univocally determines the rate matrix and the equilibrium distribution, as shown by (\ref{eq:eq_t}) and (\ref{eq:K_inf2}).

An interesting pathway for future research would be to find 
optimal recipes to infer the rate matrix $\bK$ and the equilibrium 
distribution $\bp^{\rm eq}$ from {\it partial} 
observations of the entries of matrix $\bt$, 
for more general kinetic networks, 
where MFPTs between adjacent states do not encode the full 
distribution of MFPTs.
%more general processes that Smoluchowski equation

\subsection{Constructing Transition Matrices from MFPTs}
For completeness, we show in this section how to construct
transition matrices and equilibrium distributions
from MFPTs in discrete-time dynamics. 
From (\ref{eq:zeta_disc}) and (\ref{eq:return}), one has $[\bp^{\rm eq}]^T\bt'=(1+\zeta')\bone_n^T$, where $1/(1+\zeta')=\bone_n^T \bt'^{-1}$ follows from normalization of $\bp^{\rm eq}$. Hence, $\bp^{\rm eq}$ can be computed from the matrix of MFPTs as 
\begin{equation}
    [\bp^{\rm eq}]^T=\frac{\bone_n^T \bt'^{-1}}{\bone_n^T \bt'^{-1}\cdot \bone_n}.
\end{equation}
An expression for the transition matrix $\bQ$, can be obtained by setting $\bQ=e^{\bK \tau}$ in equation (\ref{eq:mfpt_cont2}). Rewriting this
in vector notation 
\begin{equation}
    \bt'(\bI-\bQ)=\bone_n\bone_n^T-\bD_n^{-1}\bQ
\end{equation}
and rearranging for $\bQ$ gives 
\begin{equation}
    \bQ=(\bI-\bD_n\bt')^{-1}(\bp\bone_n^T-\bD_n\bt').
    \label{eq:Q_t}
\end{equation}
An alternative expression for $\bQ$ can be derived as follows. 
Starting with equation (\ref{eq:mfpt_discrete}), rewriting $\lambda'_\ell/(1-\lambda'_\ell)=1/(1-\lambda'_\ell)-1$, using the spectral representation of the identity matrix element
$I_{jk}=\sum_\ell \psi_j^{(\ell)}\phi_k^{(\ell)}$ and repeating the 
same reasoning that led to equation (\ref{eq:t_K}), we obtain 
\begin{equation}
    t'_{jk}\!=\!\frac{1}{p_j^{\rm eq}}\left[I_{jk}\!+\!(\bp^{\rm eq}\bone_n^T\!+\!\bI\!-\!\bQ)_{jj}^{-1}\!-\!(\bp^{\rm eq}\bone_n^T\!+\!\bI\!-\!\bQ)_{jk}^{-1}
    \right]
\end{equation}
Similarly to equation (\ref{eq:D_K}), this can be cast in vector notation
\begin{equation}
    \bD_n \bt=\bI+\bz'\bone_n^T-(\bp^{\rm eq} \bone_n^T+\bI-\bQ)^{-1}
    \label{eq:D_Q}
\end{equation}
where $z'_j=[(\bp^{\rm eq} \bone_n^T+\bI-\bQ)^{-1}]_{jj}$.
Rearranging for $\bQ$ and requiring $\bQ\bp^{\rm eq}=\bp^{\rm eq}$ gives 
$\bz'=\bD_n \bt \bp^{\rm eq}$ and 
\begin{equation}
    \bQ=\bI+\bp^{\rm eq}\bone_n^T-(\bI-\bD_n \bt' +\bD_n \bt' \bp^{\rm eq}\bone_n^T)^{-1}.
    \label{eq:Q_t_alt}
\end{equation}
It can be easily shown that (\ref{eq:Q_t_alt}) and (\ref{eq:Q_t}) coincide, by 
multiplying (\ref{eq:Q_t_alt}) times $(\bI-\bD_n \bt' +\bD_n \bt' \bp^{\rm eq}\bone_n^T)$ from left, expanding the products and using $\bone_n^T \bQ=\bone_n^T$.
Note that in contrast to rate matrices, the computation of transition matrices 
will in general require the inversion of two matrices, e.g. $\bt'$ and 
$\bI-\bD_n\bt'$.

\subsection{Coarse Graining Protocols that Preserve MFPTs}
Setting up this unified framework for discussing kinetic properties such as correlation functions and mean first passage times, is deeply useful for investigating new relations and interpreting the results physically. As an example we use this framework to derive a coarse graining protocol which preserves the MFPTs of the system.

Coarse graining involves projecting a high dimensional dynamics on to some coarse lower dimensional space. This involves grouping together microstates (labeled by lower case indices $i$, $j$) in to macrostates (labeled by upper case indices $I$, $J$). In what follows, we will 
denote with $P_I(t)$ the occupation probability of the macrostates 
$I=1,\ldots,N$, with $N<n$. Clearly, this must be equal to the sum of the 
probabilities of all microstates $i$ in the macrostate $I$, i.e.
$P_I(t)=\sum_{i\in I}p_i(t)$.
 
There has been much recent research in to how best to perform a kinetic coarse graining \cite{hummer2014optimal,variational,kells_mfpt}. Here we show how the link between mean first passage times and correlation functions makes it straightforward to find the coarse grained rate matrix which enforces a particular MFPT condition.

A reasonable condition to enforce would be that if we choose two (different) macrostates with equilibrium probability, then the mean first passage time between them is the same as if we choose two microstates from within the macrostates with equilibrium probability, i.e.

\begin{equation}
t_{JI}=
\frac{1}{P_I^{\rm eq}P_J^{\rm eq}}\sum_{i \in I, j \in J}p^{\rm eq}_j p^{\rm eq}_i t_{ji}
-\frac{1}{(P_J^{\rm eq})^2}\sum_{i, j \in J}p^{\rm eq}_j p^{\rm eq}_i t_{ji},
\label{eq:MFPT_CG}
\end{equation}
where the second term on the right hand side removes the contribution from microstates belonging to the same 
macrostate and 
ensures that  $t_{II}=0~\forall~I$, while $P^{\rm eq}_I=\sum_{i\in I}p^{\rm eq}_i$.

On the right hand side, we can encode the summation in to an $n\times N$ aggregation matrix $\bA$, where $A_{jJ}=1$ if $j \in J$ and is zero otherwise. On the left hand side, we can make use of (\ref{eq:t_K}) to 
express the MFPT in the coarse-grained system, in terms of the 
coarse grained rate matrix $\mathbf{R}$ and the coarse grained equilibrium probabilities $\mathbf{P}^{\rm eq}$, to get
\begin{eqnarray}
(\bA^T \dmat_n \mathbf{t} \dmat_n \bA)_{JI}\frac{1}{P_I^{\rm eq}} -
\frac{1}{P_J^{\rm eq}}(\bA^T \dmat_n \mathbf{t} \dmat_n \bA)_{JJ}\\
=(\bP^{\rm eq}\mathbf{1}_N^T-\mathbf{R})^{-1}_{JJ} 
- (\bP^{\rm eq}\mathbf{1}_N^T-\bR)^{-1}_{JI}
\nonumber\\
\end{eqnarray}
where $\bone_N$ is an $N$-dimensional vector with all the 
entries equal to $1$.
Upon defining $\bD_N$ the $N\times N$ diagonal matrix with $\mathbf{P}^{\rm eq}$ along its diagonal, we can rewrite the above as
\begin{eqnarray}\label{eq:Req}
(\bA^T \dmat_n \mathbf{t} \dmat_n \bA \bD_N^{-1})_{JI} -
(\bA^T \dmat_n \mathbf{t} \dmat_n \bA \bD_N^{-1})_{JJ}
\nonumber\\
=(\bP^{\rm eq}\mathbf{1}_N^T-\mathbf{R})^{-1}_{JJ} 
 - (\bP^{\rm eq}\mathbf{1}_N^T-\bR)^{-1}_{JI}
\nonumber\\
\end{eqnarray}
Finally, defining $u_J=(\bA^T \dmat_n \mathbf{t} \dmat_n \bA \bD_N^{-1})_{JJ}$ and  $v_J=(\bP^{\rm eq}\mathbf{1}_N^T-\mathbf{R})^{-1}_{JJ}$, equation (\ref{eq:Req}) can be written in matrix form and rearranged to yield an expression for the reduced rate matrix

\begin{equation}
    \bR = \bP^{\rm eq}\mathbf{1}_N^T - [(\mathbf{v}+\mathbf{u})\mathbf{1}_N^T- \bA^T\bD_n\mathbf{t}\bD_n\bA\bD_N^{-1}]^{-1}.
    \label{eq:R_v_u}
\end{equation}
The vector $\bv$ can be determined by demanding that $\bP^{\rm eq}$ is the steady state of the dynamics described by $\bR$, i.e. 
$\bR\bP^{\rm eq}=0$. Using $\bone_N^T \bP^{\rm eq}=1$,
$\bD_N^{-1}\bP^{\rm eq}=\bone_N$, $\bA \bone_N=\bone_n$ and 
$\bD_n \bone_n=\bp_n^{\rm eq}$, as well as that an invertible matrix 
has the same eigenvectors as its inverse (with inverse 
eigenvalues), we get 

\be
\bv= \bP^{\rm eq}-\mathbf{u}+\bA^T \bD_n \bt \bp^{\rm eq}
\label{eq:v}
\ee
Substituting (\ref{eq:v}) in 
(\ref{eq:R_v_u}) this finally gives

\begin{eqnarray}
    \bR &=&
    \bP^{\rm eq}\mathbf{1}_N^T - [\bP^{\rm eq}\mathbf{1}_N^T 
    +\bA^T \bD_n \bt \bp^{\rm eq}\bone_N^T\nonumber\\
    &&-\bA^T\bD_n\mathbf{t}\bD_n\bA \bD_N^{-1}]^{-1}.
    \label{eq:R_D}
\end{eqnarray}
%that can be further simplified using \ref{eq:zeta_def}
%\begin{eqnarray}
%    \bR &=&
%    \bP^{\rm eq}\mathbf{1}_N^T - [(1+\zeta)\bP^{\rm eq}\mathbf{1}_N^T 
%    \nonumber\\
%    &&-\bA^T\bD_n\mathbf{t}\bD_n\bA \bD_N^{-1}]^{-1}
%\end{eqnarray}
%
We check below that this automatically satisfies also the condition $\mathbf{1}_N^T \bR=0$. By multiplying the above equation times $\bone_N^T$ from left and equating to zero, we 
get 

\begin{equation}
    \zeta \bone_N^T=\bone_n^T \bD_n \bt 
    \bD_n \bA \bD_N^{-1}
\end{equation}
where we have used $\bone_N^T \bA^T=\bone_n^T$, 
$\bone_n^T \bD_n=[\bp^{\rm eq}]^T$ and
(\ref{eq:zeta_def}).
Substituting (\ref{eq:D_K}) into the above equation 

\begin{eqnarray}
\zeta \bone_N^T
     &=&\bone_n^T \bz \bone_n^T 
    \bD_n \bA \bD_N^{-1}
    \nonumber\\
    &&-\bone_n^T (\bp \bone_n^T-\bK)^{-1}
    \bD_n \bA \bD_N^{-1},
    \label{eq:identity}
\end{eqnarray}
and using $\bone_n^T \bz=(1+\zeta)\bone_N^T$,
$\bone_n^T(\bp \bone_n^T-\bK)^{-1} =\bone_n^T$, 
$\bone_n^T\bD_n \bA \bD_N^{-1}=\bone_N^T$,
and $\bone_n^T \bp=1$, shows that (\ref{eq:identity}) is 
identically satisfied.

We conclude this section by noting that, 
if information on MFPTs and equilibrium distribution
is available, the rate matrix of the 
coarse-grained system, as given in (\ref{eq:R_D}), can be 
computed at low computational cost, as it only
requires the inversion of a matrix with low dimensionality 
$N<n$. 
%Note that $\bP^{\rm eq}$ can be similarly obtained by inverting the low-dimensional matrix 
%\ref{eq:MFPT_CG} and using \ref{eq:eq_t}.

\subsection{Retrieval of Hummer-Szabo Coarse Graining}
In this section we show that the proposed coarse 
graining, based on equating MFPTs, coincides with the one
proposed
by Hummer and Szabo in \cite{hummer2014optimal},  
which equates the areas underneath the correlation functions

\be
\sum_{i\in I, j\in J}\int_0^\infty dt\,C_{ij}(t)=\int_0^\infty dt\, C_{IJ}(t)
\label{eq:integral_C}
\ee
By inserting (\ref{eq:D_K}) in (\ref{eq:R_D}), we have 

\begin{eqnarray}
    \bR &=& 
    \bP^{\rm eq}\mathbf{1}_N^T - [\bP^{\rm eq}\mathbf{1}_N^T + \bA^T \bz \bone_n^T \bp^{\rm eq} \bone_N^T
    \nonumber\\
    &&-
    \bA^T (\bp^{\rm eq}\bone_n^T\!-\!\bK)^{-1} \bp^{\rm eq} \bone_N^T
    -\bA^T\bz\bone_n^T\bD_n\bA \bD_N^{-1}
    \nonumber\\
    &&+\bA^T(\bp^{\rm eq}\bone_n^T\!-\!\bK)^{-1}\bD_n\bA \bD_N^{-1}]^{-1}
\end{eqnarray}
Using $(\bp^{\rm eq}\bone_n^T\!-\!\bK)^{-1} \bp^{\rm eq}=\bp^{\rm eq}$, $\bA^T \bp^{\rm eq}=\bP^{\rm eq}$
and $\bone_n^T \bD_n \bA \bD_N^{-1}=\bone_N^T$ this simplifies to 
\begin{eqnarray}
    \bR &=& 
    \bP^{\rm eq}\mathbf{1}_N^T - [
    \bA^T(\bp^{\rm eq}\bone_n^T\!-\!\bK)^{-1}\bD_n\bA \bD_N^{-1}]^{-1}
\end{eqnarray}
which coincides with the expression derived by Hummer-Szabo 
by imposing (\ref{eq:integral_C}). In contrast to 
(\ref{eq:R_D}), this formulation requires the inversion of 
a large dimensional matrix, hence (\ref{eq:R_D}) may be 
computationally more efficient when MFPTs and equilibrium 
distribution are known.
\\

\subsection{Variational principle for Kemeny Constant in Hummer-Szabo 
Coarse Graining}
In \cite{kells_mfpt} we have shown that a variational principle holds for the second largest eigenvalue of the rate matrix in the system coarse-grained according to the Hummer-Szabo prescription, 
namely its inverse (corresponding to the relaxation time in the coarse-grained system) is smaller than or equal to the inverse  
second largest eigenvalue of the rate matrix of the 
original system (giving the relaxation time of the 
original system). This variational principle has been used in \cite{variational} 
to identify optimal clustering protocols. In this section we show that 
a similar variational principle holds for the Kemeny constant 
itself. Summing (\ref{eq:MFPT_CG}) over $J$ and rewriting $\sum_J \sum_{j\in J}=\sum_j$
\begin{equation}
\sum_j\sum_{i \in I}p^{\rm eq}_j p^{\rm eq}_i t_{ji}
-\sum_J\frac{P_I^{\rm eq}}{P_J^{\rm eq}}\sum_{i, j \in J}p^{\rm eq}_j p^{\rm eq}_i t_{ji}
=\sum_J P^{\rm eq}_JP^{\rm eq}_It_{JI}
\end{equation}
and finally using (\ref{eq:mfpt_cont4}) we obtain
\begin{equation}
    \zeta =\sum_J\frac{1}{P_J^{\rm eq}}\sum_{i, j \in J}p^{\rm eq}_j p^{\rm eq}_i t_{ji}+
    \zeta ^{\rm CG}
\end{equation}
where $\zeta^{\rm CG}$ is the Kemeny constant in the coarse-grained system.
Since the first term on the RHS of the equation above is 
non-negative, we have 
\begin{equation}
    \zeta^{\rm CG}\leq \zeta.
\end{equation}
This extends the variational principle previously found for the relaxation time, to the sum of all the timescales in the system.
We intend to explore in future work 
variational clusterings based on Kemeny constants.

\section{Conclusions and Outlook}
In this study we have 
presented and linked together 
%revisited 
several results existing in the literature 
for mean first passage times and the Kemeny constant 
and have provided new relations in terms 
of correlation functions. These relations lead to a new 
writing of the Kemeny constant, and a simple 
interpretation in terms of decorrelation times. 

In addition, we have provided a recipe to infer the equilibrium distribution and 
the rate matrix of a process, from measurements of MFPTs. This does not require the estimation of 
committor probabilities and it only requires the inversion of a single 
matrix (with MFPTs between pairs of states as entries). 
For systems whose transitions are well approximated by 
memoryless jumps between adjacent states, as the one dimensional Smoluchowski process, MFPTs between any pair of states can be 
expressed in terms of MFPTs between adjacent states, 
hence the rate matrix can be constructed from the sole
measurements of MFPTs between adjacent states, using this recipe. 

This observation provides an intuitive 
explanation for the accuracy of milestoning techniques 
in inferring the whole MFPTs distribution, from short 
trajectories between adjacent milestones, which has been pointed out in \cite{milestoning_Szabo19, milestoning_Vanden08}.  
An interesting pathway for future work would be to 
define optimal recipes to infer rate matrices, from partial observations of MFPTs, in more complex kinetic networks, 
where MFPTs between adjacent states are not 
sufficient to reconstruct the full MFPTs matrix. 

The derived relation between rate matrices and MFPTs, given in 
equation (\ref{eq:K_inf2}), may find 
application in several domains. 
For example, in transport networks, the mean 
travelling times of passengers between two stations (a proxy for MFPTs),
may be readily available from smart cards, 
and can be used to infer the rates at which passengers 
move along the links of the network,
which might be more difficult to measure in practice. Often, a simple diffusive process (controlled by the degrees of the nodes) is assumed, but due to the varying importance of different nodes, this assumption may be 
invalid \cite{LeeHsuLin2014}. Equation (\ref{eq:K_inf2}) may thus be used to 
model such processes more accurately. 

Another application we can mention, is the inference of gene regulatory 
networks from the time series generated in gene knock-out experiments \cite{Villani}, which provide information on the first time at which the expression of a gene $j$ is modified, as a result of knocking out a gene $i$. 
This can be regarded as the MFPT to reach node $j$ from node $i$ 
on the relevant gene regulatory network. Using this information, an 
effective rate matrix can be computed via (\ref{eq:K_inf2}), which may give 
information on the rate at which a perturbation of gene $i$ propagates 
to gene $j$, thus providing insights on the interactions between genes.

Finally, we have shown how the relations between MFPTs and 
rate matrices can be used to introduce 
clustering protocols that preserve MFPTs. We have shown that the resulting expression for the coarse-grained rate 
matrix coincides with the one derived by 
Hummer-Szabo, and can be computed at low computational cost when information 
about MFPTs and equilibrium distribution in the original system is available. Finally, we have shown that such coarse-graining 
leads to a variational principle for the Kemeny constant, 
which may 
be used to optimise the coarse-graining protocol. We intend to 
investigate this in a further study.

\appendix

\section{Equivalence between (\ref{eq:correct}) and (\ref{eq:K_inferred})}
We start with equation (\ref{eq:K_inferred}) and multiply left and right hand sides times $\bt$, from left, and 
times $\bD_n$ from right
\begin{equation}
    \bt \bK \bD_n= \bt \bp^{\rm eq} \bone_n^T \bD_n -[\bD_n^{-1} \bp^{\rm eq} \bone_n^T \bt^{-1}-\bt (\bI-\bp^{\rm eq} \bone_n^T)\bt^{-1}]^{-1}
\end{equation}
Using (\ref{eq:zeta_def}) and $\bD_n^{-1}\bp^{\rm eq}=\bone_n$, we get 
\begin{eqnarray}
    \bt \bK \bD_n&=& \bt \bp^{\rm eq} [\bp^{\rm eq}]^T -\left[\frac{1}{\zeta}\bone_n [\bp^{\rm eq}]^T- \left(\bI-\frac{1}{\zeta}\bt\bp^{\rm eq} [\bp^{\rm eq}]^T\right)\right]^{-1}
    \nonumber\\
&=&\bt \bp^{\rm eq} [\bp^{\rm eq}]^T +\left[\bI -\frac{1}{\zeta}(\bone_n +\bt\bp^{\rm eq})[\bp^{\rm eq}]^T\right]^{-1}
\end{eqnarray}
Upon using the Sherman-Morrison formula
\begin{equation}
(\bI+\bu \bv^T)^{-1}=\bI-\frac{\bu \bv^T}{1+\bu^T \bv}
\end{equation}
(\ref{eq:zeta_def}) and $\bp^T \bone_n=1$, we find 
\begin{equation}
    \bt \bK \bD_n=-\bone_n \bp^T +\bI
\end{equation}
from which (\ref{eq:K_inf2}) follows.
%\nocit mae{*}

\section*{Acknowledgements}
All the authors gratefully thank Attila Szabo (NIDDK, NIH), 
for numerous discussions and suggestions. A.K. is supported 
by the EPSRC Centre for Doctoral Training in Cross-Disciplinary Approaches to
Non-Equilibrium Systems (CANES, EP/L015854/1).

\bibliography{aipsamp}% Produces the bibliography via BibTeX.

\end{document}